# Universal oscillations in counting statistics

C. Flindt[1,2,*], C. Fricke[3], F. Hohls[3], T. Novotný[2], K. Netočný[4], T. Brandes[5], R. J. Haug[3]

[1]*Department of Physics, Harvard University, 17 Oxford Street, Cambridge, MA 02138, USA.* [2]*Department of Condensed Matter Physics, Faculty of Mathematics and Physics, Charles University, Ke Karlovu 5, 12116 Prague, Czech Republic.* [3]*Institut für Festkörperphysik, Leibniz Universität Hannover, D 30167 Hannover, Germany.* [4]*Institute of Physics, Academy of Sciences of the Czech Republic, Na Slovance 2, 18221 Prague, Czech Republic.* [5]*Institut für Theoretische Physik, Technische Universität Berlin, D 10623 Berlin, Germany.*

[*]Corresponding author, email: flindt@physics.harvard.edu

**Noise is a result of stochastic processes that originate from quantum or classical sources. Higher-order cumulants of the probability distribution underlying the stochastic events are believed to contain details that characterize the correlations within a given noise source and its interaction with the environment, but they are often difficult to measure. Here we report measurements of the transient cumulants $\langle\langle n^m \rangle\rangle$ of the number $n$ of passed charges to very high orders (up to $m=15$) for electron transport through a quantum dot. For large $m$, the cumulants display striking oscillations as functions of measurement time with magnitudes that grow factorially with $m$. Using mathematical properties of high-order derivatives in the complex plane we show that the oscillations of the cumulants in fact constitute a universal phenomenon, appearing as functions of almost any parameter, including time in the transient regime. These ubiquitous oscillations and the factorial growth are system-independent and our theory provides a unified interpretation of previous theoretical studies of high-order cumulants as well as our new experimental data.**

Counting statistics concerns the probability distribution $P_n$ of the number $n$ of random events that occur during a certain time span $t$. One example is the number of electrons that tunnel through a nanoscopic system.[1-4] The first cumulant of the distribution is the mean of $n$, $\langle\langle n \rangle\rangle = \langle n \rangle$, the second is the variance, $\langle\langle n^2 \rangle\rangle = \langle n^2 \rangle - \langle n \rangle^2$, the third is the skewness, $\langle\langle n^3 \rangle\rangle = \langle (n - \langle n \rangle)^3 \rangle$. With increasing order the cumulants are expected to contain more and more detailed information on the microscopic correlations that determine the stochastic process. In general, the cumulants $\langle\langle n^m \rangle\rangle = S^{(m)}(z=0)$ are defined as the $m$-th derivative with respect to the counting field $z$ of the cumulant generating function (CGF) $S(z) = \ln \sum_n P_n e^{nz}$. Recently, theoretical studies of a number of different systems have found that the high-order cumulants oscillate as functions of certain parameters,[5-9] however, no systematic explanation of this phenomenon has so far been given. Examples include oscillations of the high-order cumulants of transport through a Mach-Zender interferometer as functions of the Aharonov-Bohm flux,[6] and in transport through a double quantum dot as functions of the energy dealignment between the two quantum dots.[8] As we shall demonstrate, oscillations of the high-order cumulants in fact constitute a universal phenomenon which is to be expected in a large class of stochastic processes, independently of the microscopic details. Inspired by recent ideas of M. V. Berry for the behavior of high-order derivates of complex functions,[10] we show that the high-order cumulants for a large variety of stochastic processes become oscillatory functions of basically any parameter, including time in the transient regime. We develop the theory underlying this surprising phenomenon and present the first experimental evidence of universal oscillations in the counting statistics of transport through a quantum dot.

We first present our experimental data. In our setup (Fig. 1a), single electrons are driven through a quantum dot and counted[11-16] using a quantum point contact. The quantum dot is operated in the Coulomb blockade regime, where only a single additional electron at a time is allowed to enter and leave. A large bias-voltage across the quantum dot ensures that the electron transport is uni-directional. Electrons enter the quantum dot from the source electrode at rate $\Gamma_S = \Gamma$ and leave via the drain electrode with rate $\Gamma_D = \Gamma(1-a)/(1+a)$, where $-1 \leq a \leq 1$ is the asymmetry parameter.[14,15] A nearby quantum point contact (QPC) is capacitively coupled to the quantum dot and used as a detector for real-time counting of the number of transferred electrons during transport: when operated at a conductance step edge, the QPC current is highly sensitive to the presence of localized electrons on the dot. By monitoring switches of the current through the QPC (see Fig. 1b) it is thus possible to detect single electrons as they tunnel through the quantum dot and thereby obtain the distribution of the number of transferred electrons $P_n$. In the experiment we fix the asymmetry parameter $a$ and monitor the time evolution of the number of passed charges. This requires a single long time-trace of duration $T$ during which a large number of tunnelling events are counted. The time-trace is divided into a large number $N$ of time segments of length $t$. From the number of electrons counted in each time segment we find the probability distribution $P_n$ from which the cumulants as functions of measurement time are obtained. In this approach $t$ can be varied continuously as it was recently shown experimentally up to the 5$^{th}$ cumulant.[15] For the experiment considered here roughly 670,000 electrons were counted during a time span of $T = 770$ s. This allowed us to estimate $P_n$ for $t$ in the range from 0 to about 2 ms. In Fig. 2 we show the corresponding experimental results for the time evolution of the high-order cumulants up to order 15. Most remarkably, the cumulants show transient oscillations as functions of time that get faster and stronger in magnitude with increasing order of the cumulants.

Together with the experimental results, we show theoretical fits of the cumulants as functions of measurement time. The model is based upon a master equation[18] and takes into account the finite bandwidth of the detector,[15,19] which is not capable of resolving very short time intervals between electron tunnelling events (see supporting information). The detector bandwidth $\Gamma_Q$ and the rates $\Gamma_D$ and $\Gamma_S$ are the only input parameters of the model and they can be extracted from the distribution of switching times, shown in Fig. 1c, between a high and a low current through the quantum point contact. The calculated cumulants show excellent agreement with the experiment for the first 15 cumulants as seen in Fig. 2.

The master equation calculation per se does not provide any hints concerning the origin and character of the oscillations seen in the experiment. Similarly, in the theoretical studies of high-order cumulants mentioned previously,[5-9] no systematic explanation of the origin of these oscillations has been proposed. Here, we point out the universal character of the oscillations of high-order cumulants and develop the underlying theory. The following analysis applies to a large class of stochastic processes, including the aforementioned theoretical studies and the experiment described above. We thus consider a general stochastic process with a CGF denoted by $S(z,\lambda)$. Here, all relevant quantities related to the system – collectively denoted as $\lambda$ – enter as real parameters. As it follows from complex analysis, the asymptotic behaviour of the high-order cumulants is determined by the analytic properties of the CGF in the complex $z$-plane. We consider the generic situation, where the CGF $S(z,\lambda)$ has a number of singularities $z_j$, $j = 1, 2, 3, ...$, in the complex plane that can be either poles or branch-points. Exceptions to this scenario do exist, e.g., the simple Poisson process, whose CGF is an entire function, i.e. it has no singularities. Such exceptions, however, are non-generic and are excluded in the following, although they can be addressed by analogous methods.[10] Close to a singularity $z \approx z_j$ the CGF takes the form $S(z,\lambda) \approx A_j/(z-z_j)^{\mu_j}$ for some $A_j$ and $\mu_j$. The corresponding derivatives for $z \approx z_j$ are $\partial_z^m S(z,\lambda) \approx (-1)^m A_j B_{m,\mu_j} /(z-z_j)^{m+\mu_j}$ with $B_{m,\mu_j} = (\mu_j + m - 1)(\mu_j + m - 2)...\mu_j$ for $m \geq 1$. The approximation of the derivatives becomes increasingly better away from $z \approx z_j$ as the order $m$ is increased. This is also known as Darboux's theorem.[10,17] For large $m$, the cumulants are thus well-approximated as a sum of contributions from all singularities,

$$\langle\langle n^m \rangle\rangle \approx \sum_j (-1)^{\mu_j} A_j B_{m,\mu_j} / z_j^{m+\mu_j} \quad (1).$$

This simple result determines the large-order asymptotics of the cumulants. Generally, the singularities come in complex-conjugate pairs, ensuring that the expression above is real. While actual calculations of the high-order cumulants using this expression may be cumbersome, the general result displays a number of ubiquitous features. In particular, we notice that the magnitude of the cumulants grows factorially with the order $m$ due to the factors $B_{m,\mu_j}$. Furthermore, writing $z_j = |z_j| e^{iArg(z_j)}$ with $|z_j|$ being the absolute value of $z_j$ and $Arg(z_j)$ the corresponding complex argument, we also see that the most significant contributions to the sum come from the singularities closest to $z=0$. The relative contributions from other singularities are suppressed with the relative distance to $z=0$ and the order $m$, such that they can be neglected for large $m$. Most importantly, we recognize that the high-order cumulants become oscillatory functions of *any* parameter among $\lambda$ that changes $Arg(z_j)$ as well as of the cumulant order $m$. This important observation shows that the high-order cumulants for a large class of CGFs will oscillate as functions of almost any parameter.

For a simple illustration of these concepts, we consider a charge transfer process described by a CGF reading

$$S(z,\lambda) = \frac{\Gamma t}{1+a}(w_{z,a} - 1) + \ln\frac{1 + q_{z,a} e^{-2w_{z,a}\Gamma t/(1+a)}}{1 + q_{z,a}} \quad (2)$$

with $w_{z,a} \equiv \sqrt{(1-a^2)e^z + a^2}$, $q_{z,a} \equiv -(1-w_{z,a})^2/(1+w_{z,a})^2$ and $\lambda = \{t, a, \Gamma\}$. This would correspond to our experiment in the case of an ideal detector with infinite bandwidth ($\Gamma_Q \to \infty$, see supporting information). Let us start by analyzing the first, linear-in-time term of the CGF which corresponds to the long-time limit. This term has branch points at $z_j = -\ln(1/a^2 - 1) + (2j+1)i\pi$, $j = \ldots -1, 0, 1, \ldots$, for which the argument of the square-root entering the definition of $w_{z,a}$ is zero. Parameters corresponding to these singularities are $A_j = i\Gamma t|a|/(1+a)$ and $\mu_j = -1/2$. Clearly, the positions of these singularities vary as the asymmetry parameter is changed, which modifies the complex argument $Arg(z_j)$, and we thus expect oscillations of the high-order cumulants in the long-time limit as functions of the asymmetry parameter $a$. In Fig. 3 we show the approximation for the cumulants obtained by including in the sum only the contributions from the singularities $z_0$ and $z_{-1}$ that are closest to $z = 0$. The approximation is compared with exact calculations of the cumulants obtained by direct differentiation of the CGF in the long-time limit and shows nearly perfect agreement for large orders. The 5$^{th}$ cumulant in the long-time limit has already been measured as a function of the asymmetry parameter, showing some indication of the onset of these oscillations.[15]

Finally, we now return to our experimental data presented in Fig. 2. In the transient regime we see that the high-order cumulants oscillate as functions of measurement time. This is due to the time dependence of the dominating singularities of the CGF. In the case of an ideal detector ($\Gamma_Q \to \infty$, see supporting information) the CGF at finite times in Eq. (2) has time-dependent singularities when the argument of the logarithm is zero. These singularities have the form $z_{k,j} = x_k + (2j+1)i\pi$, $k = 1, 2, \ldots$, $j = \ldots -1, 0, 1, \ldots$, where $x_k \equiv \ln[(a^2 + u_k^2)/(1-a^2)]$ and $u_k$ solves the transcendental equation $2u_k\Gamma t/(1+a) - 4\arctan(1/u_k) = 2\pi(k-1)$ (see Fig. 4a). The derivatives of the logarithmic singularity encountered here can be treated with our theory by formally setting $\mu_{k,j} = 0$ and $\mu_{k,j}A_{k,j} = 1$. In Fig. 4b we see that the approximation for the 15$^{th}$ cumulant as function of time, using the two time-dependent singularities $z_{1,-1}$ and $z_{1,0}$ closest to $z = 0$ for the given time interval, agrees well with exact calculations in the limit of an ideal detector ($\Gamma_Q \to \infty$), taking $a = -0.34$ as in the experiment. The curves are also in good agreement with the experimental results in Fig. 2, showing that the oscillations cannot be dismissed as an experimental artefact due to, e.g., the finite bandwidth of the detector. Of course, in the long-time limit the cumulants relax to their linear-in-time asymptotics given by the first term of the CGF. The low-order cumulants ($m = 4-7$) seen in Fig. 2, normalized with respect to the first cumulant, clearly reach their long-time limits for $t \geq 1.5$ ms. This does not contradict the fact that the cumulants oscillate as functions of time in a given finite time interval for high enough order.

The experimental and theoretical results presented in this work clearly demonstrate the universal character of the oscillations of high-order cumulants. In our experiment the high-order cumulants oscillate as functions of time in the transient regime. As our theory shows, such oscillations are however predicted to occur as functions of almost any parameter in a wide range of stochastic processes, regardless of the involved microscopic mechanisms. The universality of the oscillations stems from general mathematical properties of cumulant generating functions: as some parameter is varied, dominating singularities move in the complex plane, causing the oscillations. Oscillations of high-order cumulants have been seen also in other branches of physics, including quantum optics[20] and elementary particle physics,[21] further demonstrating the universality of the phenomenon.

**Methods**

**Device.** The quantum dot and the quantum point contact were fabricated using local anodic oxidation techniques with an atomic force microscope on the surface of a GaAs/AlGaAs heterostructure with electron density $n = 4.6 \times 10^{15} \, m^{-2}$ and mobility $\mu = 64 \, m^2/Vs$. With this technique the two-dimensional electron gas residing 34 nm below the heterostructure surface is depleted underneath the oxidized lines on the surface. A number of in-plane gates were also defined, allowing for electrostatic tuning of the quantum point contact and electrostatic control of the tunnelling barriers between the quantum dot and the source and drain electrodes.

**Measurement.** The experiment was performed at an electron temperature of about 380 mK, as determined from the width of thermally broadened Coulomb blockade resonances. To avoid tunnelling from the drain to the source contact of the quantum dot due to thermal fluctuations, we applied a bias of 330 µV across the quantum dot. The QPC detector was tuned to the edge of the first conduction step. The current through the QPC was measured with a sampling frequency of 100 kHz. This sufficiently exceeded the bandwidth of our experimental setup of about 40 kHz. The tunnelling events were extracted from the QPC signal using a step detection algorithm.

**Error estimates.** To estimate the error of the experimentally determined cumulants we created an ensemble of simulated data using the same rates as observed in the experiment. We then extracted the cumulants for each simulated data set in the ensemble and determined the ensemble variance of the cumulants for each order $m$ as function of time $t$. The error bars in Fig. 2 show the square-root of the variance.

**Acknowledgements** We thank N. Ubbelohde (Hannover, Germany) for his support in the development of the data analysis algorithms. W. Wegscheider (Regensburg, Germany) provided the wafer and B. Harke (Hannover, Germany) fabricated the device. The work was supported by the Villum Kann Rasmussen Foundation (C. Fl.), the Czech Science Foundation 202/07/J051 (C. Fl., T. N., K. N.), Federal Ministry of Education and Research of Germany via nanoQUIT (C. Fr., F. H., R. J. H.), German Excellence Initiative via the "Centre for Quantum Engineering and Space-Time Research" (C. Fr., F. H., R. J. H.), research plan MSM0021620834 financed by the Ministry of Education of the Czech Republic (T. N.), project AV0Z10100520 in the Academy of Sciences of the Czech Republic (K.N.), and Deutsche Forschungsgemeinschaft Project BR 1528/5–1 (T. B.).

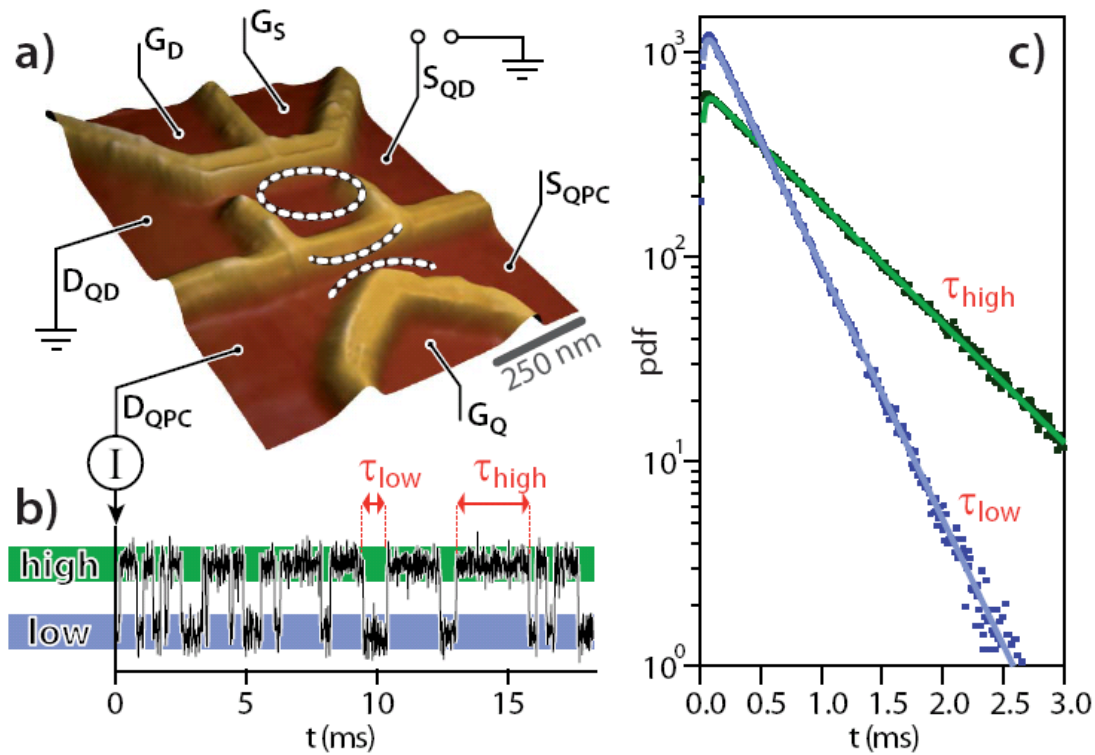

**Figure 1. Real-time counting of electrons tunnelling through a quantum dot. a)** Atomic force microscope topography of the quantum dot (QD, dashed ring) and the quantum point contact (QPC, dashed lines). The gates $G_S$ and $G_D$ are used to electrostatically control the tunnelling barriers between QD and source ($S_{QD}$) and drain ($D_{QD}$) electrodes, respectively. The gate $G_Q$ is used to tune the QPC. **b)** Typical time-trace of the current from $S_{QPC}$ to $D_{QPC}$ via the QPC. The current switches between a high and a low level: as an electron tunnels onto the QD from $S_{QD}$, the QPC current is suppressed. The suppression is lifted as the electron leaves via $D_{QD}$. Electrons passing through the QD are counted by monitoring switches of the QPC current. **c)** Measured (unnormalized) probability density functions (pdf) for the switching times $\tau_{low}$ and $\tau_{high}$ shown with dots. For short times the data display a kink due to the finite detector bandwidth. The lines show theoretical predictions taking $\Gamma_D = 2.97\,\text{kHz}$, $\Gamma_S = 1.46\,\text{kHz}$, and $\Gamma_Q = 40\,\text{kHz}$ as fitting parameters. The asymmetry parameter is $a = -0.34$.

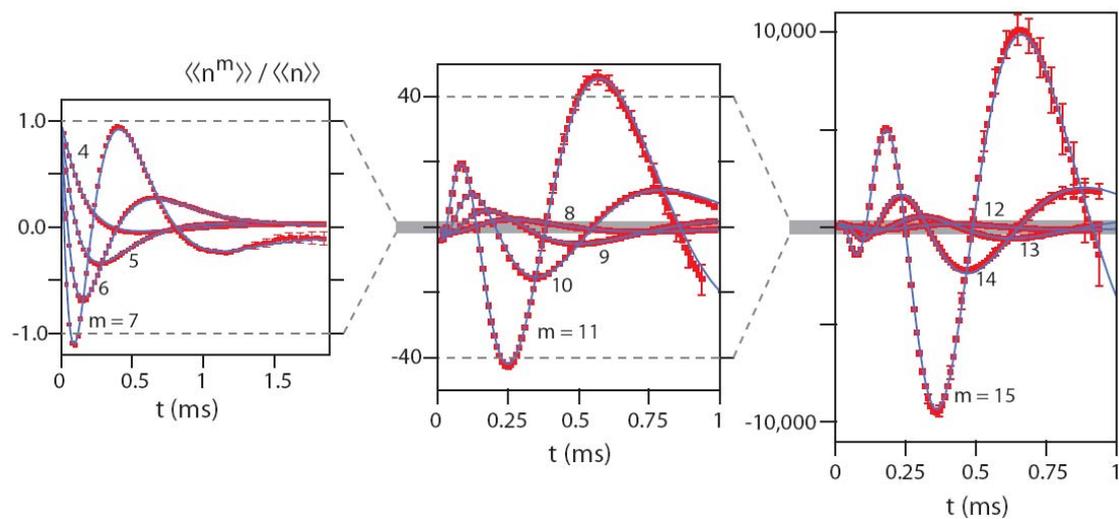

**Figure 2. Measurement of high-order cumulants.** Experimental results (squares) for the time evolution of the first 4-15 cumulants. The cumulants clearly show oscillations as functions of time, with increasing magnitude and number of oscillations for higher orders $m$. The theoretical model (full lines) shows excellent agreement with the experimental data. For very high orders ($m \geq 10$) the finite number of electron counts during the experiment limits the statistical accuracy to times shorter than approximately 1 ms. The estimates for the error bars are discussed in the Methods section. Parameters corresponding to the data shown here are given in the caption of Fig. 1.

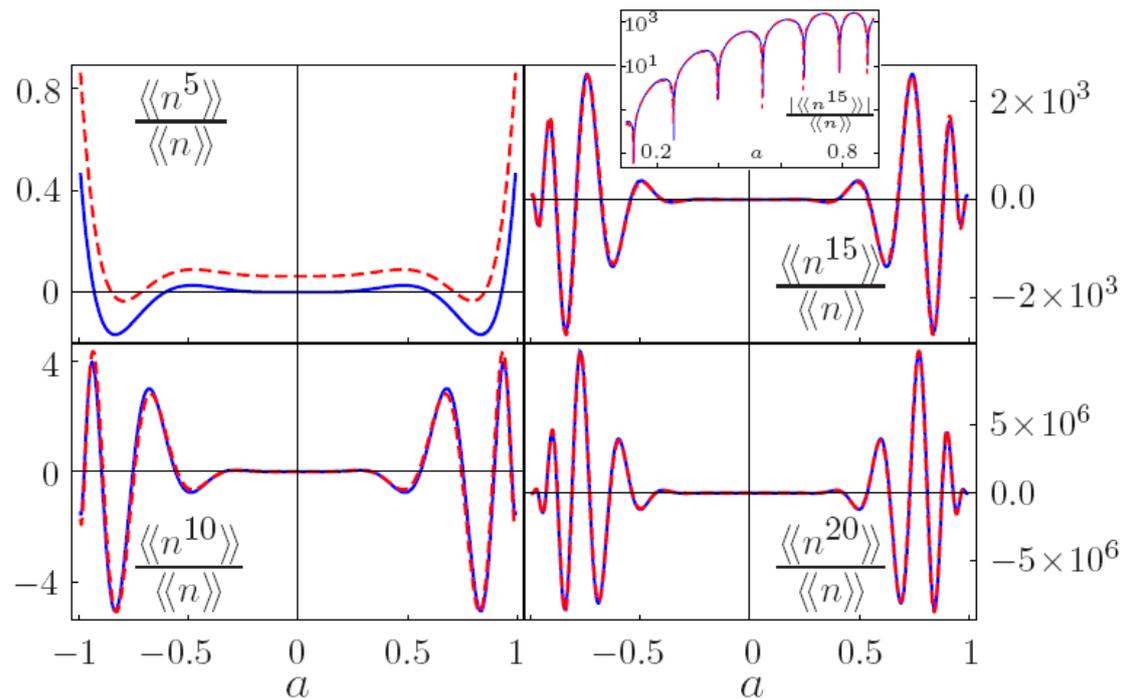

**Figure 3. Universal oscillations of cumulants.** High-order cumulants in the long-time limit as functions of the asymmetry parameter $a$. We assume an ideal detector with infinite bandwidth. Full lines correspond to exact theoretical results for the cumulants, while dashed lines show the asymptotic approximation using the two dominating singularities $z_0$ and $z_{-1}$ closest to $z = 0$ (see Fig. 4a). As the order $m$ is increased, the asymptotic approximation becomes better, and for $m \geq 10$, the two curves are nearly indistinguishable. The cumulants are clearly oscillatory functions of the asymmetry parameter. As the order is increased, the number and amplitudes of the oscillations grow. The inset shows the absolute value of the 15[th] cumulant on a logarithmic scale for $0.1 < a < 0.9$.

414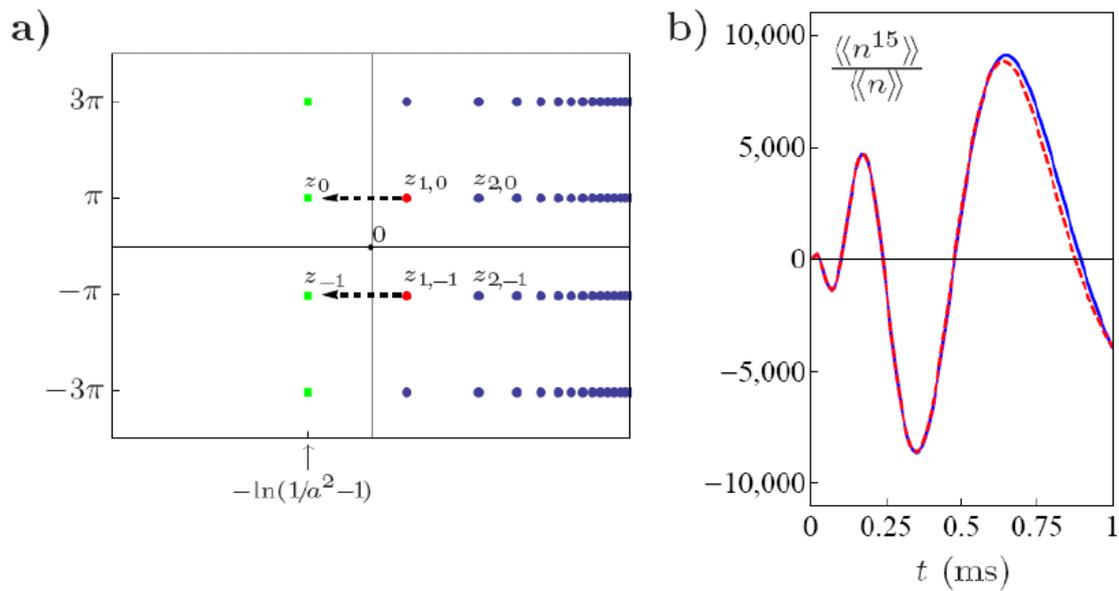

**Figure 4. Singularities in the complex plane and universal oscillations as function of measurement time. a)** Complex plane with the singularities of the CGF. The singularities $z_j = -\ln(1/a^2 - 1) + (2j+1)i\pi$, $j = \ldots -1, 0, 1, \ldots$, corresponding to the linear-in-time term of the CGF are shown with green squares. Among these singularities, $z_0$ and $z_{-1}$ are closest to $z = 0$ and thus responsible for the oscillations of the cumulants in the long-time limit seen in Fig. 3. The time-dependent singularities $z_{k,j}$, $k = 1, 2, \ldots$, $j = \ldots -1, 0, 1, \ldots$, are shown with colored circles. The arrows indicate the motion with time of the two dominating singularities $z_{1,-1}$ and $z_{1,0}$ (shown with red). Here, parameters are $a = -0.34$, $t = 0.3$ ms, and $\Gamma_Q \to \infty$. **b)** Transient oscillations of the 15$^{\text{th}}$ cumulant as function of time with $a = -0.34$ and $\Gamma_Q \to \infty$. The full line corresponds to exact theoretical results, while the dashed line shows the asymptotic approximation using the dominating singularities $z_{1,-1}$ and $z_{1,0}$. For $t \geq 0.6$ ms a slight deviation is seen. We attribute this to the singularites $z_{2,-1}$ and $z_{2,0}$, which also come close to 0. The curves agree well with the experimental results in Fig. 2, even if the finite-bandwidth of the detector has not been included here.

# Supporting information
# "Universal oscillations in counting statistics"

C. Flindt, C. Fricke, F. Hohls, T. Novotný, K. Netočný, T. Brandes, and R. J. Haug

## I. MODELING THE QD-QPC SYSTEM

To describe our experiment we use a model previously employed in Refs. [1–3]. We thus consider a master equation for the probability vector $|p(n,t)\rangle = [p_{00}(n,t), p_{10}(n,t), p_{01}(n,t), p_{11}(n,t)]^T$, containing the probabilities $p_{kl}(n,t)$ for the quantum dot (QD) to be in a charge state with $k = 0, 1$ extra electrons, and the quantum point contact (QPC) indicating that $l = 0, 1$ extra electrons reside on the QD, while $n$ electrons have been counted during the time span $t$. The probability distribution for the number of counted electrons is equal to the sum of the four probabilities. This can be written as the inner product of the probability vector with the vector $\langle 1| = [1, 1, 1, 1]$, i.e., $P(n,t) = \langle 1|p(n,t)\rangle$. The corresponding moment generating function is defined as $\hat{P}(z,t) \equiv \sum_n P(n,t) e^{nz} = \langle 1|\hat{p}(z,t)\rangle$ with $|\hat{p}(z,t)\rangle \equiv \sum_n |p(n,t)\rangle e^{nz}$ and $z$ being the counting field. The dynamics of $|\hat{p}(z,t)\rangle$ is governed by a master equation (see e.g. Ref. [4]) of the form $\frac{d}{dt}|\hat{p}(z,t)\rangle = M(z)|\hat{p}(z,t)\rangle$ with solution $|\hat{p}(z,t)\rangle = e^{M(z)t}|\hat{p}(z,0)\rangle$, where

$$M(z) = \begin{pmatrix} -\Gamma_S & \Gamma_D & \Gamma_Q e^z & 0 \\ \Gamma_S & -(\Gamma_D + \Gamma_Q) & 0 & 0 \\ 0 & 0 & -(\Gamma_S + \Gamma_Q) & \Gamma_D \\ 0 & \Gamma_Q & \Gamma_S & -\Gamma_D \end{pmatrix}. \quad (1)$$

The unresolved (with respect to the number of passed electrons) probability distribution is $\hat{P}(0,t) = \sum_n P(n,t)$, whose time evolution is determined by $M(0)$. Using $M(0)$ we can also calculate the waiting time distributions [5] for switches of the QPC current using the methods recently developed in Ref. [6]. For the probability of switching from a low to a high QPC current with a time difference $\tau$ we find

$$P_{\text{low}}(\tau) = \frac{\Gamma_Q \Gamma_D\, e^{-(\Gamma_S + \Gamma_D + \Gamma_Q)\tau/2}}{\sqrt{(\Gamma_S + \Gamma_D + \Gamma_Q)^2 - 4\Gamma_Q \Gamma_D}} \left( e^{\tau\sqrt{(\Gamma_S+\Gamma_D+\Gamma_Q)^2 - 4\Gamma_Q\Gamma_D}/2} - e^{-\tau\sqrt{(\Gamma_S+\Gamma_D+\Gamma_Q)^2 - 4\Gamma_Q\Gamma_D}/2} \right). \quad (2)$$

For an ideal detector ($\Gamma_Q \to \infty$) the standard exponential law $P_{\text{low}}(\tau) = \Gamma_D e^{-\Gamma_D \tau}$ is recovered. The corresponding distribution for switches from a low to a high QPC current $P_{\text{high}}(\tau)$ is identical to $P_{\text{low}}(\tau)$ when $\Gamma_S$ and $\Gamma_D$ are interchanged. The theoretical predictions for the switching times are compared with experimental results, allowing us to extract the rates $\Gamma_S$, $\Gamma_D$, and $\Gamma_Q$ (see Fig. 1).

## II. CUMULANT GENERATING FUNCTION

The cumulant generating function (CGF) is obtained as $S(z, \lambda) = \ln \hat{P}(z, t) = \ln \langle 1 | e^{M(z) t} | p(z, 0) \rangle$. The variable $\lambda$ denotes all system parameters including the time $t$. We assume that the system has reached the stationary state when counting begins, i.e., we take as the initial condition $|\hat{p}(z, 0)\rangle = |p^{\text{stat}}\rangle$, the (unique) normalized solution to $M(0)|p^{\text{stat}}\rangle = 0$. Whereas $\hat{P}(z, t)$ is an entire function, $S(z, \lambda)$ has logarithmic singularities at the zeros of $\hat{P}(z, t)$ in the complex-$z$ plane for any finite time $t$. However, a different non-analytic structure can still emerge in the long-time limit. For example, the CGF corresponding to an ideal detector ($\Gamma_Q \to \infty$) reads

$$S(z, \lambda) = \frac{\Gamma t}{1+a} (w_{z,a} - 1) + \ln \frac{1 + q_{z,a} e^{-\frac{2 w_{z,a}}{1+a} \Gamma t}}{1 + q_{z,a}} \tag{3}$$

where $\Gamma \equiv \Gamma_S$,

$$w_{z,a} \equiv \sqrt{(1-a^2) e^z + a^2}, \qquad q_{z,a} \equiv -\left(\frac{1 - w_{z,a}}{1 + w_{z,a}}\right)^2, \tag{4}$$

and $a = (\Gamma_S - \Gamma_D)/(\Gamma_S + \Gamma_D)$ is the asymmetry parameter. The CGF can be decomposed in three terms $S(z, \lambda) = A(z) t + B(z) + C(z, t)$ in which $A(z)$ yields the limit of $S(z, \lambda)/t$ for $t \to \infty$ and $C(z, t)$ accounts for finite-time corrections, which are exponentially suppressed at long times. Clearly, $A(z)$ has branch points where the argument of the square-root is zero. However, at finite times, these branch point singularities are regularized by the presence of the finite-time correction $C(z, t)$, ensuring that only the logarithmic singularities of $C(z, t)$ remain [7]. In the long-time limit, the logarithmic singularities accumulate and produce the branch point singularities of $A(z)$. In this limit, the asymmetry parameter $a$ is the only relevant variable of the CGF. In contrast, at finite times, the analytic structure of the CGF is time-dependent in a non-trivial manner as discussed in the following section.

The cumulants of the distribution are determined by the derivatives of the CGF $S(z, \lambda)$ with respect to the counting field $z$ at the origin. In practice, it may be difficult to calculate very high orders of these derivatives by direct differentiation, since the resulting expressions become very large. Instead, we determine the derivatives at $z = 0$ using a Cauchy integral, writing $S(z, \lambda) = \frac{1}{2\pi i} \oint_C dz' \frac{S(z', \lambda)}{z' - z}$, where $C$ is a positively oriented contour surrounding $z' = z$. The cumulants are then

$$\langle\langle n^m \rangle\rangle(\lambda) = S^{(m)}(0, \lambda) = \frac{m!}{2\pi i} \oint_C dz' \frac{S(z', \lambda)}{z'^{m+1}}. \tag{5}$$

Deforming the contour into a circle with radius $\varepsilon$, small enough that no singularities of $S$ are enclosed, we may parametrize it as $C : z = \varepsilon\, e^{i\theta}$, $\theta \in [-\pi, \pi]$. The cumulants can then be written as

$$\langle\langle n^m \rangle\rangle(\lambda) = S^{(m)}(0, \lambda) = \frac{m!}{2\pi} \int_{-\pi}^{\pi} d\theta\, e^{-im\theta} \frac{S(\varepsilon\, e^{i\theta}, \lambda)}{\varepsilon^m} \tag{6}$$

where the integral can be performed numerically, even for moderately large orders of $m$. While this approach suffices for the given problem, more sophisticated methods are also available for more complicated cases [8]. The model calculations shown in Fig. 2 were obtained by evaluating numerically the expression in Eq. (6) using the full expression for the CGF in the case of a finite detector bandwidth.

## III. TRANSIENT OSCILLATIONS

For times that are not much larger than the relaxation time $1/(\Gamma_S + \Gamma_D)$, time enters as a relevant parameter in the CGF $S(z, \lambda)$ and the corresponding singularities are in general time-dependent. This is the reason for the oscillatory dependence of the cumulants on time. In the long-time limit, the dependence of the singularities on the asymmetry parameter is relatively simple. At finite times, the dependence of the singularities on time and the asymmetry parameter is more complicated as we shall see.

Since $S(z, \lambda)$ is periodic by construction, $S(z + 2\pi i, \lambda) = S(z, \lambda)$, it is sufficient to restrict ourselves to the strip $-\pi \leq \Im z \leq \pi$ containing the origin. The transient part of the CGF, denoted $C(z, t)$ in the previous section, has time-dependent logarithmic singularities that coincide with the zeroes of the function $Q(z, \lambda) = 1 + q_{z,a} \exp(-\frac{2w_{z,a}}{1+a} \Gamma t)$, see Eq. (3). They all lie on the line $\Im z = \pi$ and one finds that $Q(x + \pi i, \lambda) > 0$, whenever $x < x_0 = \ln[a^2/(1-a^2)]$. Therefore, all the singularities are localized rightwards from the dominating singularity of the non-transient part $z_0 = x_0 + \pi i$ (see Fig. 4a). Considering now $x > x_0$, we get

$$Q(x + \pi i, \lambda) = 1 - \exp i\left(4 \arctan \frac{1}{u} - \frac{2u}{1+a} \Gamma t\right) \tag{7}$$

with $u = \sqrt{(1-a^2)e^x - a^2}$, and thus find infinitely many (time-dependent) zeroes at $z_{k,0}(t) = \ln[(a^2 + u_k^2)/(1-a^2)] + \pi i$, by solving the equations

$$\frac{2u_k}{1+a}\Gamma t - 4\arctan\frac{1}{u_k} = 2\pi(k-1), \quad k = 1, 2, 3, \ldots \quad (8)$$

The resulting picture now depends on whether the asymmetry parameter $a$ is larger or smaller than the critical value $a^* = 1/\sqrt{2}$ for which $x_0 = 0$. In the high-asymmetry regime, $|a| > a^*$, the dominating time-dependent singularity is $z_{1,0}(t)$, and its complex conjugate $\overline{z_{1,0}(t)} = z_{1,-1}(t)$ (see Fig. 4a). The CGF can then be written $S(z, \lambda) = S_{\text{reg}}(z, \lambda) + \ln[z - z_{1,0}(t)][z - z_{1,-1}(t)]$ with $S_{\text{reg}}(z, \lambda)$ being non-singular around $z_{1,0}(t)$ and $z_{1,-1}(t)$. For a fixed finite time interval and in the large cumulant-order asymptotics, other time-dependent singularities become suppressed and the logarithmic singularities at $z_{1,0}(t)$, $z_{1,-1}(t)$ become responsible for the emergence of cumulant oscillations.

The low-asymmetry regime, $|a| < 1/\sqrt{2}$, corresponding to the actual experimental setup, leads to a somewhat more involved analytical structure, since the logarithmic singularities now may have a negative real part. The dominating role is then played by the singularities on the line $\Im z = \pi$ (and on the line $\Im z = -\pi$) as they move by the origin (see Fig. 4a) with time. For the experimentally relevant time span, the oscillations are governed by the first pair of singularities $z_{1,0}(t)$, $z_{1,-1}(t)$. For longer times, a rather complicated oscillation pattern is expected due to the interference with the next pairs of singularities that come close to the origin.

The long-time asymptotics of any fixed cumulant does not follow simply from the above finite-time analysis: as time increases the gap separating the dominating singularities from the others approaches zero and, in fact, the logarithmic singularities tend to form a continuous spectrum. Consequently, the smallest cumulant order for which our finite-time theory is meaningful also increases with time. However, in the limit of $t$ going to infinity, we know the limit of $S(z, \lambda)/t$ explicitly as it is given by $A(z)$; its branch point singularities then determine the cumulant asymptotics. Apparently, we consider here two different asymptotic regimes, namely, the transient and the stationary ones, which correspond to the limits of large cumulant order and of large time taken in the opposite order: for the former, the large order asymptotics of the cumulants is considered first, whereas, for the latter, the first limit is that of large time. The different structure of the finite-time and of the stationary singularities then explains why the two limits are not interchangeable. Naturally, in interpreting the time dependence of a moderate-order cumulant, one has to take into account that it exhibits a non-trivial crossover between both asymptotic regimes.

---